\documentclass{aastex}
\input{epsf}
\usepackage{emulateapj5,apjfonts}

\def\oiii{[{\sc O\, iii}]}

\def\feka{Fe K$\alpha$}
\def\chandra{{\it Chandra}} 
 
\def\asca{{\it ASCA}} 
\def\hst{{\it HST}} 
 
\def\sax{{\it BeppoSAX}} 
 
\def\rosat{{\it ROSAT}}

\def\lum{erg s$^{-1}$}
\def\flux{erg cm$^{-2}$ s$^{-1}$}
\def\nh{cm$^{-2}$}
\def\arcsec{$^{\prime\prime}$}

\def\cir{Circinus}

\def\ltsima{$\; \buildrel < \over \sim \;$}
\def\simlt{\lower.5ex\hbox{\ltsima}} % < over ~
\def\gtsima{$\; \buildrel > \over \sim \;$}
\def\simgt{\lower.5ex\hbox{\gtsima}} % > over ~

\slugcomment{Accepted for publication in the Astrophysical Journal Letters}

\shorttitle{CHANDRA IMAGING OBSERVATIONS OF CIRCINUS}

\shortauthors{SAMBRUNA ET AL.}

\begin{document}

\title{X-ray imaging of the Seyfert 2 galaxy Circinus with \chandra} 

\author{Rita M. Sambruna,\altaffilmark{1} 
W. N. Brandt,\altaffilmark{1}
G. Chartas,\altaffilmark{1}
Hagai Netzer,\altaffilmark{2} 
S. Kaspi,\altaffilmark{1}
G. P. Garmire,\altaffilmark{1}
John A. Nousek,\altaffilmark{1}
and
K. A. Weaver\altaffilmark{3}
}
\altaffiltext{1}{Department of Astronomy and Astrophysics, 525 Davey
Laboratory, The Pennsylvania State University, University Park, PA 16802.}
\altaffiltext{2}{School of Physics and Astronomy, Raymond and Beverly
Sackler Faculty of Exact Sciences, Tel-Aviv University, Tel-Aviv 69978, Israel.}
\altaffiltext{3}{Laboratory for High Energy Astrophysics, Code 660,
NASA/Goddard Space Flight Center, Greenbelt, MD 20771.}

\begin{abstract} 

We present results from the zeroth-order imaging of a \chandra\ HETGS
observation of the nearby Seyfert 2 galaxy \cir. Twelve X-ray sources
were detected in the ACIS-S image of the galaxy, embedded in diffuse 
X-ray emission. The latter shows a prominent ($\sim$ 18\arcsec) soft
``plume'' in the N-W direction, coincident with the \oiii\ ionization
cone. The radial profiles of the brightest X-ray source at various
energies are consistent with an unresolved (FWHM $\sim$ 0.8\arcsec)
component, which we identify as the active nucleus, plus two extended
components with FWHMs $\sim$ 2.3\arcsec\ and 18\arcsec,
respectively. In a radius of 3\arcsec, the nucleus contributes roughly
the same flux as the extended components at the softest energies
(\simlt 2 keV). However, at harder energies ($>$ 2 keV), the
contribution of the nucleus is dominant. The zeroth-order ACIS
spectrum of the nucleus exhibits emission lines at both soft and hard
X-rays, including a prominent \feka\ line at 6.4 keV, showing that
most of the X-ray lines previously detected with \asca\ originate in a
compact region (\simlt 15 pc). Based on its X-ray spectrum, we argue
that the 2.3\arcsec\ extended component is scattered nuclear radiation
from nearby ionized gas. The large-scale extended component includes
the emission from the N-W plume and possibly from the outer starburst
ring.

\end{abstract}


\section{Introduction}

Observations of Seyfert 2 galaxies with \rosat\ and \asca\ showed that
many distinct components contribute to the soft X-ray radiation from
these sources. Emission from diffuse gas, extending up to a few kpc
around the nucleus, was imaged with \rosat\ (e.g., Morse et al. 1995;
Weaver et al. 1995; Matt et al. 1994), and interpreted as gas
associated to a starburst or scattered nuclear radiation (Wilson et
al. 1992; Matt et al. 1994). The \asca\ spectra of Seyfert 2s are
consistent with obscured hard X-ray continua and often show emission
lines at soft and hard X-rays (e.g., Turner et al. 1997), of unknown
origin.  Attempts to disentangle the contributions of the nuclear and
extranuclear components were hampered by the poor angular resolution of
the \asca\ detectors.  The excellent spatial resolution
(0.5\arcsec/pixel) and spectral capabilities of \chandra\ make it
uniquely suited to this task.

Here we study the case of the \cir\ galaxy. At a distance of 4 Mpc,
\cir\ is a well-studied Seyfert 2 with spectacular manifestations of
nuclear and extranuclear activity, including an \oiii\ ionization cone
(Marconi et al. 1994) and two starburst rings at
%a luminous and rapidly
%variable water maser (Greenhill et al. 1997), a bright IR/optical
%nucleus with a rich host of coronal lines (Oliva et al. 1994; Moorwood
%et al. 1996), polarized radio lobes (Elmouttie et al. 1995), an O[III]
%ionization cone (Marconi et al. 1994), and two starburst rings at
$\sim$ 2\arcsec\ and 10\arcsec\ from the nucleus (Wilson et al. 2000
and references therein). In X-rays, imaging with \rosat\ showed an
unresolved X-ray source at the nuclear position and three discrete
X-ray sources within 1\arcmin\ (Guainazzi et al. 1999). \asca\
observations show a rich emission line spectrum (Sako et al. 2000a;
Netzer, Turner, \& George 1998; Matt et al. 1996), attributed to the
active nucleus. We performed a 60 ks \chandra\ HETGS observation of
\cir\ in an effort to determine the origin of its X-ray emission; here
we present the first results from an analysis of the zeroth-order ACIS
image, while in a companion paper we discuss the HETGS spectrum. At
the distance of the source, 1\arcsec=19 pc.
%Throughout this paper, we assume $H_0=75$ km$^{-1}$ s$^{-1}$
%Mpc$^{-1}$ and $q_0$=0.5, thus at the distance of the galaxy,
%1\arcsec\ $\sim$ 30 pc.

\section{Observations and Data Analysis}

The \cir\ galaxy was observed with the High Energy Transmission
Grating Spectrometer (HETGS; Canizares et al. 2000, in prep) on 2000
June 6, with ACIS-S (Garmire et al. 2000) in the focal plane. The
observation was continuous, with the target at the aimpoint of the S3
chip. In order to reduce pileup in the zeroth-order, a customized
subarray window of 600 rows was used, yielding a frametime of 2.1 s.
With this choice, pileup is negligible: the nucleus (the brightest
X-ray point source in the field) has $\sim$ 0.2 counts/frame
corresponding to a pileup fraction of 3\% (see the \chandra\
Proposer's Observatory Guide, Figure 6.25). The data were reduced using
the \verb+CIAO+ v.1.1.5 software provided by the \chandra\ X-ray
Center (CXC). Gain correction was applied using updated calibration
files appropriate for the observing epoch, and we filtered for times
of bad aspect and ``bad'' pixels. Only \asca\ grades 0, 2, 4, and 6
were accepted. The total net exposure time was 60,223 s. We checked
that the S3 background was stable during the observation. ACIS spectra
were analyzed using spectral responses generated with \verb+CIAO+ for
nodes 0 and 1.

The astrometric accuracy of our \chandra\ image is $\sim$ 2.5\arcsec,
derived comparing the position of the nucleus in the ACIS image with
published values from radio and optical observations (Elmouttie et
al. 1995; Gardner \& Whiteoak 1982). This is poorer than the nominal
$\sim$ 1\arcsec\ accuracy. Unfortunately, it was not possible to
improve the absolute astrometry, as no unrelated foreground or
background sources were detected in the field of view.

%ACIS-S spectra were extracted for the point sources with 100 counts or
%more for each node, rebinned to a minimum of 20 counts per bin to
%validate the use of $\chi^2$ statistics, and analyzed within
%\verb+XSPEC+ v.11.0.1.  Spectral responses were generated using the
%\verb+CIAO+ software using calibration files appropriate for nodes 0
%and 1.

\section{Results}

The zeroth-order image of \cir\ in the energy range 0.5--8 keV is
shown in Figure \ref{figzero}.  A bright X-ray source is apparent,
which we identify with the nucleus, together with several point

%\begin{figure} 
\centerline{\epsfxsize=8.9cm\epsfbox{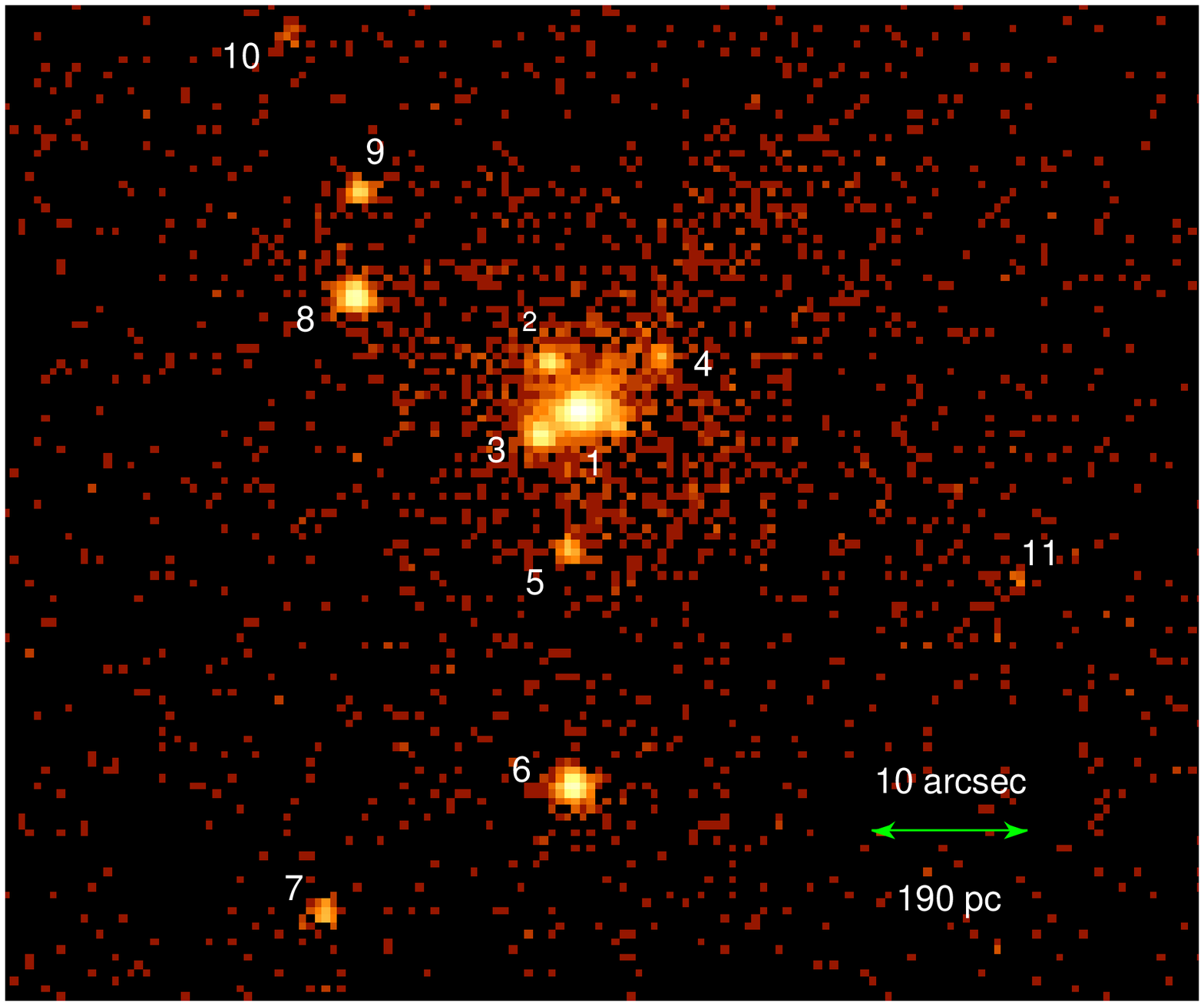}}
\centerline{}
\figcaption{\chandra\ ACIS-S image of \cir\ in 0.5--8 keV from a 60 ks
HETGS exposure. North is up and East is to the left. Twelve point
sources are detected, as numbered (see Table 2); source \# 12 is out
of the image. The nucleus is the brightest X-ray source in the
field (\# 1). Extended emission is apparent, with an asymmetric
``plume'' in a N-W direction.
\label{figzero} }
\centerline{}
\centerline{}
%\end{figure}

\noindent sources, of which at least four are located within 5\arcsec\ of the
nucleus. Using the algorithm \verb+wavdetect+ (Freeman et al. 2000)
with a significance threshold of $10^{-7}$, twelve sources are
detected in Figure \ref{figzero}.  The X-ray sources are embedded in a
faint, diffuse X-ray emission with a ``plume'' extending $\sim$
20\arcsec\ in the N-W direction.  We estimate that, within a circle of
radius 43\arcsec, $\sim$ 50\% of the 0.5--8 keV counts are from the
nuclear region, while the discrete X-ray sources contribute $\sim$
30\% of the counts and the extended emission 20\%. The total 0.5--8
keV count rate from ACIS-S in this aperture is 0.19 counts s$^{-1}$
consistent with \asca\ and \sax\ (Guainazzi et al. 1999; Matt et
al. 1996).

%We now discuss the main features of Figure 1 in some detail. A more
%thorough discussion of the observational results will be presented in
%a future publication.

\subsection{Nuclear vs. Extended X-ray emission} 

The brightest X-ray source in the field, with a total of 5,309 counts
in the 0.5--8 keV band, has X-ray coordinates $\alpha$(2000)=14h 13m
09.83s, $\delta$(2000)=--65$^{\circ}$ 20\arcmin\ 21.35\arcsec.  We
identify this source with the nuclear region of \cir.

We calculated the 0.5--8 keV radial profile of this source integrating
the counts in annuli with widths of 0.1\arcsec\ for radii \simlt
2\arcsec, and with widths 0.5\arcsec\ for radii \simgt 2\arcsec. The
profiles were background-subtracted using background counts from a
blank region in the field, and the serendipitous sources excised using
circles of radii 1.5\arcsec. The observed 0.5--8 keV profile can be
fit with three Gaussians: an unresolved point source with
FWHM=0.8\arcsec\ $\pm$ 0.09\arcsec, an extended component with
FWHM=2.3\arcsec\ $\pm$ 0.17\arcsec, and a second diffuse component
with FWHM=18\arcsec\ $\pm$ 2\arcsec. We identify the unresolved
component with the active nucleus. Based on the X-ray spectrum (see
below), the 2.5\arcsec\ component is most likely scattered nuclear
radiation by ionized gas. Its size ($\sim$ 50 pc) is consistent with
the size of the scattering mirror in Seyfert 2s from \hst\
spectropolarimetry (e.g., Kraemer, Ruiz, \& Crenshaw 1998).  The
large-scale component comprises the N-W plume, scattered radiation,
and possibly emission from the outer starburst ring. By integrating
the best-fit Gaussian models, we estimate that, in a radius of
3\arcsec, the point source contributes 70\% of the 0.5--8 keV counts,
with the inner extended \ component \ contributing most 

%\begin{figure} 
%\centerline
\hglue-1.9cm{\epsfxsize=11.2cm\epsfbox{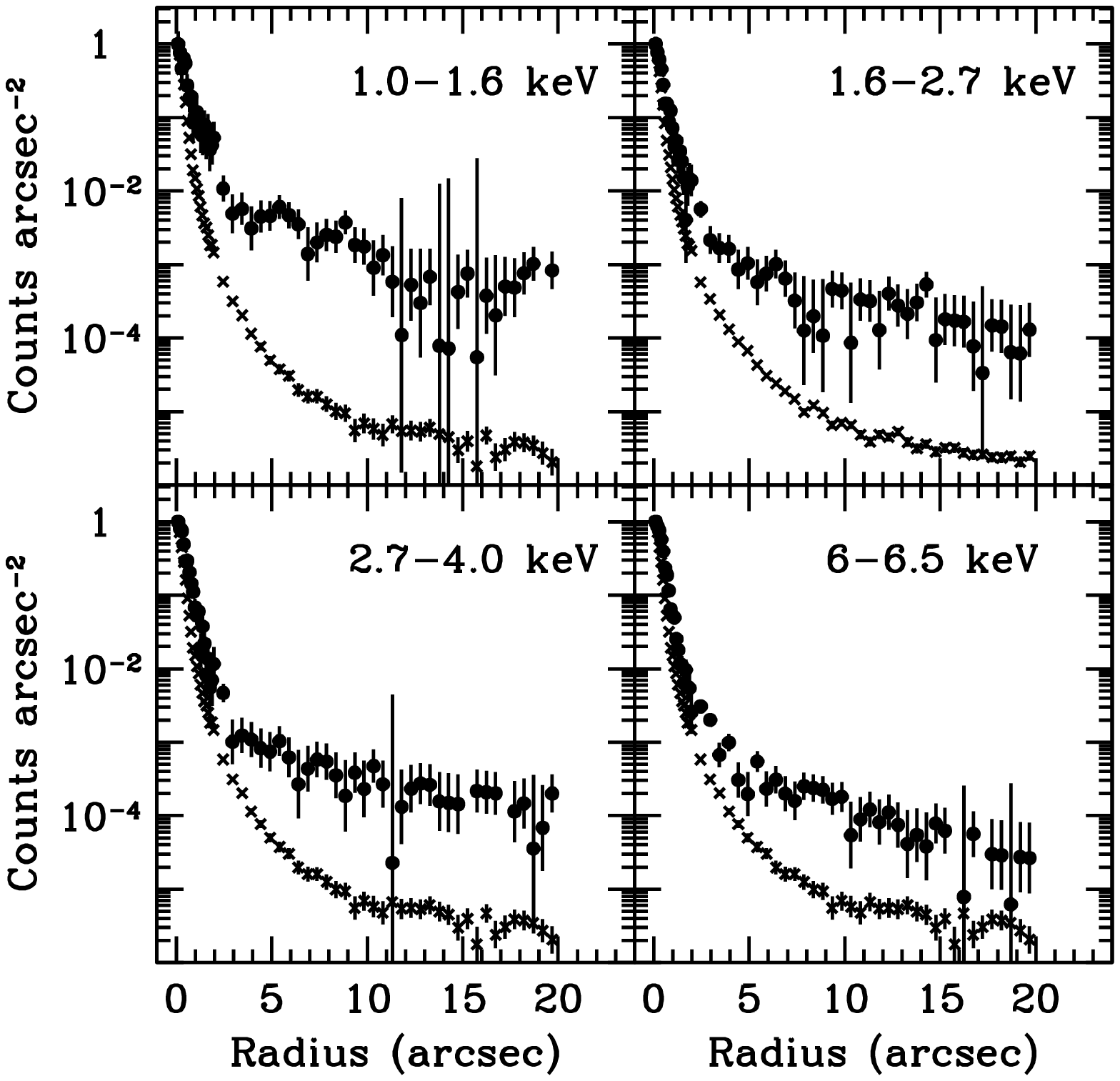}}
\vglue-2.2cm
\figcaption{Radial profiles of the nuclear region of \cir\ (source \# 1
in Table 2) in different energy bands, after excising the
serendipitous X-ray sources. Filled circles: observed
profiles. Asterisks: simulated PSF using MARX (see text). The profiles
are consistent with an unresolved (FWHM $\sim$ 0.8\arcsec) component,
plus two extended components with FWHM $\sim$ 2\arcsec\ and
20\arcsec.
\label{profiles}}
\centerline{}
\centerline{}
%\end{figure}

\noindent (29\%) of the
remaining counts.
%, the 2.3\arcsec\ diffuse emission with the inner starburst
%ring, and the large-scale extended emission with the diffuse emission
%visible in the zeroth-order image (Figure \ref{figzero}). 
%in 0.5--8 keV, the point source contains 60\% of the counts
%in an aperture of 20\arcsec, while the inner extended component
%contains 29\%.

Energy-dependent radial profiles were also computed in the 1.0--1.6,
1.6--2.7, 2.7--4.0, and 6.0--6.5 keV bands. These energy bands bracket
the ranges of emission from ionized Mg, Si, S, and from neutral Fe,
respectively, while containing at the same time enough photons to
allow meaningful signal-to-noise ratios. The profiles are shown in
Figure \ref{profiles}a--d together with the corresponding simulated
PSFs. The latter were calculated with the \verb+MARX+ simulator
(v.3.0), assuming a Gaussian spatial model with width
$\sigma$=0.17\arcsec\ to take into account aspect effects, and as
input spectral model a power law (photon index $\Gamma=-0.8$) plus an
\feka\ line, which to a first approximation describes the zero-order
ACIS spectrum. As in the case of the 0.5--8 keV profile, the
energy-dependent profiles in Figure \ref{profiles} can be fitted with
three Gaussians, an unresolved (FWHM $\sim$ 0.8\arcsec) point source,
and two extended components with FWHMs $\sim$ 2.5\arcsec\ and
14--21\arcsec, with uncertainties of $\sim$ 34\% or better. The
relative contributions of the three components at various energies and
extraction radii are reported in Table 1. They were obtained by
integrating the best-fit Gaussians of the three components.
%In a radius of 3\arcsec, the unresolved component contributes 37\%,
%57\%, 70\% and 81\% of the total counts in 1.0--1.6 keV, 1.6--2.7 keV,
%2.7--4.5 keV, and 6.0--6.5 keV, respectively, with the remaining
%counts being due to the 2.5\arcsec\ extended component.  In an
%aperture of 20\arcsec, the nucleus contributes 20\%, 43\%, 55\%, and
%72\% of the total counts in 1.0--1.6 keV, 1.6--2.7 keV, 2.7--4.5 keV,
%and 6.0--6.5 keV, respectively; the corresponding contribution from
%the inner extended component is 37\%, 33\%, 30\%, and 18\%, and the
%remaining counts are due to the large-scale extended component.
In the smaller aperture, the nucleus and the 2.5\arcsec\ component
give roughly similar contributions at soft energies (\simlt 2 keV),
while at hard X-rays (\simgt 2 keV) the nucleus becomes dominant.
\cir\ is thus a different case from the Seyfert 2 Mrk3, where recent
\chandra\ observations show that {\it all} the soft X-ray flux
originates from an extended region (Sako et al. 2000b).

\subsection{Zeroth-order ACIS spectra} 

The ACIS spectra of the nucleus in a radius 0.8\arcsec, and of the
inner extended component, between 0.8\arcsec\ and 3\arcsec, are shown
in Figure \ref{spectra}. A total of 4132 and 1214 counts were
collected for the nucleus and the extended component, respectively.
The nucleus exhibits several emission lines at both soft and hard
X-rays, including a prominent (EW $\sim$ 2.5 keV) \feka\ line. The
extended

\footnotesize 
\begin{center}
{\sc TABLE 1\\
Contributions of Nucleus and Extended Components}
\vskip 4pt
\begin{tabular}{rrrr}
\hline
\hline
{Energy}&
{Nucleus}&
{Ext. 2.5\arcsec} &
{Ext. 18\arcsec} \\
\hline
\multicolumn{4}{c}{Radius=0.8\arcsec} \\
\hline
1.0--1.6 keV & 0.72 & 0.28 & 0.00 \\
1.6--2.7 keV & 0.85 & 0.15 & 0.00 \\
2.7--4.5 keV & 0.90 & 0.10 & 0.00 \\
6.0--6.5 keV & 0.95 & 0.05 & 0.00 \\
\hline
\multicolumn{4}{c}{Radius=3\arcsec} \\
\hline
1.0--1.6 keV & 0.37 & 0.58 & 0.05 \\
1.6--2.7 keV & 0.57 & 0.42 & 0.01 \\
2.7--4.5 keV & 0.66 & 0.33 & 0.01 \\
6.0--6.5 keV & 0.81 & 0.18 & 0.01 \\
\hline
\multicolumn{4}{c}{Radius=20\arcsec} \\
\hline
1.0--1.6 keV & 0.20 & 0.37 & 0.43 \\
1.6--2.7 keV & 0.43 & 0.33 & 0.24 \\
2.7--4.5 keV & 0.55 & 0.29 & 0.16 \\
6.0--6.5 keV & 0.72 & 0.18 & 0.10 \\
\hline
\end{tabular}
%\tablenotetext{a}{Extended 1 is the 2.5\arcsec\ extended component;
%Extended 2 is the 20\arcsec\ extended component.}
\end{center}
\setcounter{table}{1}
\normalsize

\noindent component shows at least 5 emission lines at
E$_1$=0.96$^{+0.30}_{-0.18}$ keV, $E_2$=1.25 $\pm$ 0.02 keV,
$E_3$=1.78 $\pm$ 0.05 keV, $E_4$=2.25 $\pm$ 0.05 keV, and $E_5$=6.4
$\pm$ 0.02 keV, and a continuum typical of a combination of highly
ionized material (positive Gamma at low energies) together with a low
ionization reflector (negative Gamma at high energies).
%and a very flat power law continuum (photon index
%$\Gamma \sim -0.3$) between 2.3 and 6.0 keV. 
These properties suggest that a likely origin of the 2.5\arcsec\
extended component is scattering of the nuclear radiation from ionized
gas. A more detailed spectral analysis will be given in our companion
paper. Here for future reference we quote the line fluxes of the
extended component: N$_1=(4.0^{+61}_{-1.4}) \times 10^{-4}$ ph
cm$^{-2}$ s$^{-1}$, N$_2=(3.0^{+12}_{-2.2}) \times 10^{-5}$ ph
cm$^{-2}$ s$^{-1}$, N$_3=(1.3^{+1.4}_{-0.5}) \times 10^{-5}$ ph
cm$^{-2}$ s$^{-1}$, N$_4=(5.7^{+3.6}_{-1.7}) \times 10^{-6}$ ph
cm$^{-2}$ s$^{-1}$, and N$_5=(4.4^{+1.1}_{-0.7}) \times 10^{-5}$ ph
cm$^{-2}$ s$^{-1}$.
%The power law observed fluxes are F$_{0.2-2~keV} \sim 3 \times
%10^{-14}$ and F$_{2-10~keV} \sim 2 \times 10^{-12}$ \flux. The
%intrinsic luminosities are L$_{0.2-2~keV} \sim 6 \times 10^{37}$ and
%L$_{2-10~keV} \sim 3.3 \times 10^{39}$ \lum.

%\subsection{\feka\ emission around the nucleus} 

%Figure \ref{profiles}d shows a radial profile centered on the nucleus
%in the energy range 6.0--6.5 keV, compared to a point source profile,
%after removing the extranuclear point sources.  Interestingly, faint
%extended emission is present, extending at least up to $\sim$
%10\arcsec\ (300 pc). We checked for possible instrumental effects
%(e.g., scattering of the nucleus hard photons) by extracting 6.0--6.5
%keV images of two Galactic sources of intensity similar or larger than
%the \cir\ nucleus, ..... ({\bf GORDON, NAMES?})  which were observed
%with the HETG as well. (No halos larger than 5\arcsec\ were
%observed. Note also that scattering effects are included in the
%\verb+MARX+ simulation of the point-like profile in Figure
%\ref{profiles}d.) To the best of our knowledge, no other calibration
%effects are known which could produce the extended emission in Figure
%\ref{profiles}d. We thus conclude that the 6.0--6.5 keV halo is
%probably real.

%The most plausible interpretation is that the 6.0--6.5 keV represents
%fluorescent \feka\ emission from relatively cold gas around the
%nucleus. This ``\feka\ reverberation'' is similar to the one observed
%in the center of our Galaxy with \asca\ (ref...?) and is observed here
%%or the first time in an extragalactic source.

\subsection{The N-W ``plume''} 

A faint, diffuse X-ray emission is apparent in Figure \ref{figzero},
with an asymmetrical total extension in a N-W direction for $\sim$
20\arcsec\ ($\sim$ 0.4 kpc). The ``plume'' is offset by an angle of
$\sim$ 10 degrees from the dispersion direction, and coincides with
the large-scale \oiii\ emission from an archival \hst\ image (Wilson
et al. 2000).

% flux in 0.2-2kev: 8.5e-14; intrinsic lum in 0.2-2 kev: 1.4e39 

The ACIS spectrum of the plume contains only 200 counts, mostly from
0.6--2 keV.
%A very flat continuum is also present between 2 and 7 keV,
%possibly scattered radiation from the nucleus.  
While deeper ACIS observations are needed to study in detail the
origin of the X-ray emission from the \oiii\ cone, we briefly note
that a possibility suggested by the X-ray spectrum is a blend of soft
X-ray lines around 0.6--1 keV (O, Ne, Fe L), possibly from gas
photoionized from the nucleus. However, assuming an ionization
parameter $U_X \sim 0.02$ and linear extent of the gas 20\arcsec=0.6
kpc, we find that the luminosity required to ionize the gas is L$_X
\sim 8 \times 10^{43}$ \lum, more than one order of magnitude higher
than observed for the nucleus (Matt et al. 1999, and our companion
paper).  Alternatively, this component may well be due to starburst
regions at that location with a typical temperature $kT \sim 0.3$
keV. However, the signal-to-noise ratio of the data is not adequate to
test this idea.
%Alternatively, the soft X-rays may be due to thermal emission from gas
%with temperature $kT \sim 0.2$ keV. This is consistent with the values
%predicted by photoionization shock models (Dopita \& Sutherland 1995)
%for shock velocities $\sim$ 200 km s$^{-1}$, consistent with the FWHM
%of the \oiii\ filaments (Veilleux \& Bland-Hawthorn 1997).

%Above 2 keV, The ACIS spectrum is consistent with a flat component,
%described by either a flat power law ($\Gamma \sim 0.7$) or a very hot
%bremmstrahlung ($kT \sim 100$ keV).  The observed flux and intrinsic
%luminosity of the soft component are s $10^{-13}$ \flux\ and
%L$^{intr}_{2-10~keV} \sim 5 \times 10^{38}$ \lum.  

\subsection{Discrete X-ray point sources} 

We comment briefly on the properties of the serendipitous X-ray
sources, leaving more details to a future publication. Table 2 lists
the basic X-ray properties of the detected X-ray sources, including
crude spectral information in the form of hardness ratios (HR), and
the sources' intrinsic 0.5--8 keV luminosities, which are in the range
L$_X \sim 10^{37-39}$ \lum.
%including the X-ray positions (col. 2), the source net counts in
%0.2--8 keV (col. 2), the hardness ratios (col. 3), when enough counts
%are present (defined as the ratio of the counts in 2--8 keV to the
%counts in 0.5--2 keV), the observed flux (col. 4) and intrinsic
%luminosity (col. 5) in 0.2--8 keV, computed from the best-fit model to
%the ACIS spectrum unless otherwise indicated. From Table 2 it is
%apparent that the intrinsic X-ray luminosities span a narrow range
%around $10^{38-39}$ \lum.
The sources do not have obvious optical counterparts on an archival
\hst\ WFPC2 F606W image within 4\arcsec\ from \ the \ X-ray \ positions, \
down to a

%\begin{figure} 
\centerline{\epsfxsize=9.0cm\epsfbox{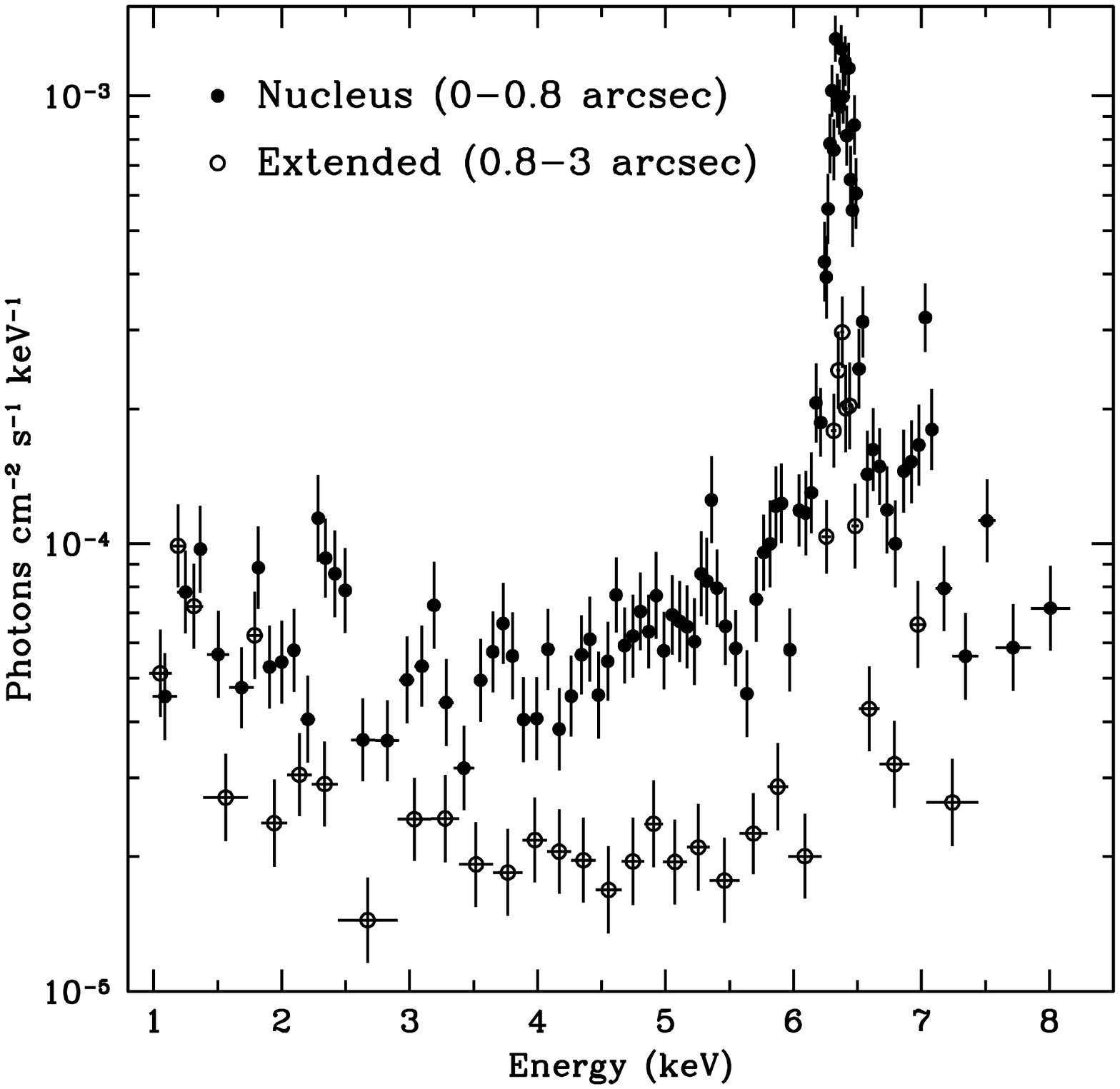}}
\figcaption{Zeroth-order ACIS spectra of the nucleus in a circle of
radius 0.8\arcsec, and of the extended component between 0.8\arcsec\
and 3\arcsec. Only spectra for node 0, where 65\% of the counts were 
collected, are presented for clarity. The spectrum of the nucleus
shows several emission lines at both soft and hard X-rays. The
spectrum of the extended component shows emission lines between 1.0
and 2.2 keV, a prominent \feka\ line at 6.4 keV, and a very flat power
law continuum between 2.5 and 6 keV. These properties suggest it is
scattered nuclear radiation from ionized gas.
\label{spectra}}
\centerline{}
\centerline{}
%\end{figure}

\noindent
limiting magnitude of $V \sim 25$ (Carollo et al. 1997). The
implied X-ray-to-optical flux ratios, $f_x/f_V$ \simgt 15.8, are
larger than for stars, normal galaxies, and AGN (Maccacaro et
al. 1988).  The ACIS spectra of the serendipitous sources exhibit
spectral cutoffs below $\sim$ 0.8--1 keV, consistent with column
densities N$_H=(5-10) \times 10^{21}$ \nh. These columns are
consistent within their large uncertainties with the Galactic value in
the direction to \cir, N$_H=(3.3 \pm 0.3) \times 10^{21}$ (Freeman et
al. 1977).
%Moreover, for Galactic gas-to-dust ratios, the X-ray column
%densities imply extinctions $A_V$=2--5, well in the range of internal
%reddening of the galaxy (Storchi-Bergmann et al. 1999).
The X-ray spectra are generally hard (HR $>$ 2, Table 2), and
inspection of the spectra shows that strong emission lines (EW $\sim$
a few hundred eV) are present in the brightest sources at both soft
and hard X-rays, indicating emission from ionized Ne, Si, S, Arg, and
Fe. Short-term ($\sim$ 1--3 hours) flux variability is also observed
in several cases; remarkably, source \# 8 shows periods of total flux
occultation every $\sim$ 20 ks. This source was also detected with
\rosat\ (Guainazzi et al. 1999) with a 0.1--2.4 keV count rate a
factor 1.6 lower than measured with \chandra. Based on these
properties, likely origins for the the serendipitous X-ray sources are
X-ray binaries and/or ultra-luminous SNRs (e.g., Chu, Chen, \& Lai
1999).

\section{Conclusions} 

A \chandra\ X-ray observation of \cir\ shows complex X-ray
emission. Besides the unresolved nucleus, several bright X-ray sources
(most likely X-ray binaries or ultra-luminous SNRs associated with the
galaxy itself) are detected within several arcsec from the nucleus, as
well as diffuse ionized/cold gas, on scales as large as 20\arcsec\
(400 pc), contributing to various degrees to the X-ray emission. We
find that at the softer energies, the X-ray emission is equally
contributed to by the unresolved nucleus (i.e., \simlt 0.8\arcsec\ or
15 pc) / and by an extended  emission on

\begin{table*}[t]
\footnotesize 
\caption{X-ray sources in the ACIS-S3 Image of Circinus 
\label{tab2}}
\begin{center}
\begin{tabular}{rllcrrrl}
\hline
\hline
{Source \#} &
{$\alpha$(2000)} &
{$\delta$(2000)} &
{0.5--8 keV Counts} &
{HR\tablenotemark{a}} &
{F$^{obs}_{0.5-8~keV}$\tablenotemark{b}} &
{L$^{intr}_{0.5-8~keV}$\tablenotemark{b}} &
{Notes} \\ [0.1cm]
\hline
1 & 14 13 09.8 & --65 20 21.4 & 5309 $\pm$ 75 & 11.7 & 11.0 & 21.0 & Nucleus \\
2 & 14 13 10.2 & --65 20 18.3 & 246 $\pm$ 18  & 2.3  & 0.3  &  2.8 & Lines at 2.3 and 1.2 keV\\
3 & 14 13 10.2 & --65 20 23.0 & 613 $\pm$ 27 & 7.4  &  1.0  & 1.9 & Variable \\
4 & 14 13 9.1  & --65 20 18.0 & 67 $\pm$ 11  & 2.3  & 1.3\tablenotemark{c} & 0.03\tablenotemark{c} & \\
5 & 14 13 9.9  & --65 20 30.0 & 121 $\pm$ 12  & 8.9  &  0.2  & 1.0 & \\
6 & 14 13 9.9  & --65 20 45.0 & 1154 $\pm$ 34 & 5.0  &  1.3 & 3.4 & Lines at 2.6 and 6.9 keV \\ 
7 & 14 13 12.4 & --65 20 53.0 & 105 $\pm$ 11  & 3.1  &  0.1 & 0.3 & Variable \\
8 & 14 13 12.1 & --65 20 14.1 & 989 $\pm$ 32 & 27.0 &  0.8 & 3.0 & Lines at 2.19, 3.03, and 4.35 keV\\
  &            &             &                &     &     & & Variable (periodic) \\ 
9 & 14 13 12.1 & --65 20 7.7  & 169 $\pm$ 14  & 6.5  &  0.2 & 0.4 & Variable \\ 
10 &14 13 12.8 & --65 19 57.7 & 21 $\pm$ 5  & $\cdots$ & 0.03\tablenotemark{c} & 0.07\tablenotemark{c} & \\
11 &14 13 5.5  & --65 20 32.0 & 21 $\pm$ 5 & $\cdots$  & 0.03\tablenotemark{c} & 0.07\tablenotemark{c} &\\
12 &14 13 3.0  & --65 20 43.5 & 19 $\pm$ 5 & $\cdots$  & 0.03\tablenotemark{c} & 0.07\tablenotemark{c} &\\
\hline
\end{tabular}
\vskip 2pt
\parbox{6in}{ % use this to define the width of the notes under the table
\small\baselineskip 9pt
\footnotesize
\indent
$\rm ^a${Hardness ratios, i.e., ratios of the 2--8 keV counts
to the 0.5--2 keV counts.} \\
$\rm ^b${Observed fluxes in $10^{-12}$ \flux, and intrinsic (i.e.,
absorption-corrected) luminosities in $10^{39}$ \lum, from the
best-fit model to the ACIS spectra.} \\
%{Columns explanation: 1=Source number (see Fig. 1a); 
%2 and 3=Source positions from \chandra; 
%4=Source net counts in 0.2--8 keV; 
%5=Hardness ratios (ratio of the 2--8 keV to 0.5--2 keV counts); 
%6=Observed flux in 0.2--10 keV, in 10$^{-12}$ \flux;
%7=Intrinsic luminosity in 0.2--10 keV in 10$^{39}$ \lum.}
$\rm ^c${Calculated assuming a power law with N$_H=1 \times
10^{22}$ \nh\ and photon index $\Gamma=1.5$, consistent with the HR.} 
}
\end{center}
\end{table*}
\normalsize

\noindent
scales $\sim$ 2\arcsec\ or 38
pc. At harder X-rays (\simgt 2 keV), the nucleus becomes
dominant. This is in contrast to Mrk3, where recent \chandra\ HETGS 
observations show that {\it all} the soft X-ray flux originate from a
resolved extended component (Sako et al. 2000b).

%This hints at a variety of intrinsic conditions in Seyfert 2s, which
%can be addressed by future \chandra\ observations.

\acknowledgements 

We are grateful to the ACIS and HETGS teams, who made these
observations possible, and to Eric Feigelson for a critical reading of
the manuscript and assistance. We ackowledge the financial support of
NASA grant NAS8--38252 (RMS; GPG PI), NASA LTSA grant NAG5--8107 (WNB,
SK), and of the Israel Science Foundation (HN).


\begin{thebibliography}{}

%\bibitem[cccc]{vvvv} Baganoff, F., et~al. 2000, ApJ, in  press

\bibitem[cccc]{vvvv} Chu, Y.-H., Chen, R., \& Lai, S.-P. 1999, in
proceedings of the STScI 1999 May Symposium ``The Largest Explosions
Since the Big Bang: Supernovae and Gamma Ray Bursts'', in press
(astro-ph/9909091) 

\bibitem[xxx (1999)]{x1bb} Carollo, C.M., Stiavelli, M., de Zeeuw,
P.T., \& Mack, J. 1997, AJ, 114, 2366 

%\bibitem[xxx (1999)]{x1bb} Dickey, J.M. \& Lockman, F.J. 1990, ARA\&A,
%28, 215 

%\bibitem[xxx (1999)]{x1bb} Dopita, M. A. \& Sutherland, R.S. 1995,
%ApJ, 455, 468 

\bibitem[xxx (1999)]{x0} Elmouttie, M., Haynes, R.F., Jones, K.L.,
Ehle, M., Beck, R. \& Wielebinski, R. 1995, MNRAS, 275, L53 

\bibitem[xxx (1999)]{x0b} Freeman, P. E., Kashyap, V., Rosner, R. \&
Lamb, D. Q.  2000, ApJ, subm.

\bibitem[xxx (1999)]{x0a} Freeman K.C., Karlsson, B., Lynga, G.,
Burrell, J.F., van Woerden, H., Goss, W.M., Mebold, U. 1977, A\&A, 55,
445 

\bibitem[xxx (1999)]{x1bc} Gardner, F.F. \& Whiteoak, J.B. 1982,
MNRAS, 201, 13p

\bibitem[xxx (1999)]{x1b} Garmire, G. et al. 2000, ApJS, subm.  

%\bibitem[xxx (1999)]{x1} Greenhill, L.J., Ellingsen, S.P., Norris,
%R.P., Gough, R.G., Sinclair, M.W., Moran, J.M. \& Mushotzky,
%R.F. 1997, ApJ, 474, L103 

%\bibitem[xxx (1999)]{x1c} Griffiths, R. et al. 2000, Science, in press  

\bibitem[xxx (1999)]{x1a} Guainazzi, M. et al. 1999, MNRAS, 310, 10 

%\bibitem[yyy (1990)]{x1cv} Koyama, K., Maeda, Y., Sonobe, T.,
%Takeshima, T., Tanaka, Y.\ and Yamauchi, S.\ 1996 PASJ, 48, 249

\bibitem[yyy (1990)]{x1cv1} Kraemer, S. B, Ruiz, J. R., \& Crenshaw,
D. M. 1998, ApJ, 508, 232 

\bibitem[xxx (1999)]{x1a1} Maccacaro, T., Gioia, I.M., Wolter, A.,
Zamorani, G. \& Stocke, J.T. 1988, ApJ, 326, 680 

\bibitem[xxx (1999)]{x1a2} Marconi, A., Morwood, A.F.M., Origlia, L. \&
Oliva, E. 1994, ESO Messenger, 78, 20  

\bibitem[xxx (1999)]{x22} Matt, G. et al. 1999, A\&A, 341, L39 

\bibitem[xxx (1999)]{x2} Matt, G. et al. 1996, MNRAS, 281, L69 

\bibitem[xxx (1999)]{x12} Matt, G., Piro, L., Antonelli, L.A., Fink,
H.H., Meurs, E.J.A. \& Perola, G.C. 1994, A\&A, 292, L13 

\bibitem[xxx (1999)]{x22a} Morse, J.A., Wilson, A.S., Elvis, M., \&
Weaver, K.A. 1995, ApJ, 439, 121 

%\bibitem[xxx (1999)]{x22} Morwood, A.F.M., Lutz, D., Oliva, E.,
%Marconi, A., Netzer, H., Genzel, R. Sturm, E. \& de Graauw, T. 1996,
%A\&A, 315, L109  

\bibitem[xxx (1999)]{x22x} Mulchaey, J.S., Colbert, E., Wilson, A.S.,
Mushotzky, R.F., \& Weaver, K.A. 1993, ApJ, 414, 144 

\bibitem[xxx (1999)]{x22y} Netzer, H., Turner, T.J., \& George,
I.M. 1998, ApJ, 504, 680 

%\bibitem[xxx (1999)]{x22x} Netzer, H. \& Turner, T.J. 1997, ApJ, 488, 694

%\bibitem[xxx (1999)]{x2a} Oliva, E., Salvati, M. Morwood, A.F. \&
%Marconi, A. 1994, A\&A, 288, 457 
 
\bibitem[xxx (1999)]{x3} Sako, M., Kahn, S.M., Paerels, F., \&
Liedahl, D.A. 2000a, ApJ, in press (astro-ph/0006146)

\bibitem[xxx (1999)]{x3b} Sako, M. et al. 2000b, ApJ Letters, in
press (astro-ph/0009323) 

%\bibitem[xxx (1999)]{x3c} Storchi-Bergmann, T., Winge, C., Ward,
%M.J. \& Wilson, A.S. 1999, MNRAS, 304, 35 

\bibitem[xxx (1999)]{x3d} Turner, T.J., George, I.M., Nandra, K. \&
Mushotzky, R.F. 1997, ApJS, 113, 23 

%\bibitem[xxx (1999)]{x3a} Veilleux, S. \& Bland-Hawthorn, J. 1997,
%ApJ, 479, L105  

\bibitem[xxx (1999)]{x4} Wilson, A. et al. 2000, AJ,
subm. (astro-ph/0006147)  

\bibitem[xxx (1999)]{x5} Wilson, A.S., Elvis, M., Lawrence, A., \&
Bland-Hawthorn, J. 1992, ApJ, 391, L75

\bibitem[xxx (1999)]{x5a} Weaver, K.A., Mushotzky, R.F., Serlemitsos,
P.J., Wilson, A.S., Elvis, M., \& Briel, U. 1995, ApJ, 442, 597 


\end{thebibliography}
\end{document}